# Image Steganography using Karhunen-Loève Transform and Least Bit Substitution

Ankit Chadha, Neha Satam, Rakshak Sood, Dattatray Bade
Department of Electronics and Telecommunication
Vidyalankar Institute of Technology
Mumbai, India

## ABSTRACT
As communication channels are increasing in number, reliability of faithful communication is reducing. Hacking and tempering of data are two major issues for which security should be provided by channel. This raises the importance of steganography. In this paper, a novel method to encode the message information inside a carrier image has been described. It uses Karhunen-Loève Transform for compression of data and Least Bit Substitution for data encryption. Compression removes redundancy and thus also provides encoding to a level. It is taken further by means of Least Bit Substitution. The algorithm used for this purpose uses pixel matrix which serves as a best tool to work on. Three different sets of images were used with three different numbers of bits to be substituted by message information. The experimental results show that algorithm is time efficient and provides high data capacity. Further, it can decrypt the original data effectively. Parameters such as carrier error and message error were calculated for each set and were compared for performance analysis.

## 1. INTRODUCTION
With the advent of various communication techniques, its applications are also on the rising scale. They include browsing, E-mailing, file transfer and remote login. Though the first three need not have utmost level security, remote login does include some kind of security breach. Number of eavesdroppers has increased since communication mediums are also employed to convey highly sensitive and confidential information. Information prone to such attacks can be secured by encoding it using various practices, mainly by adding redundant data. This gave rise to branch of cryptography namely steganography. It has been proposed as a new, alternative method to enforce intellectual property rights and protect digital media from tampering. The methods used in steganography are characterized as imperceptible, robust and secure communication of data related to the host signal, which includes embedding into and extraction from the host signal. In historic times when data was encoded, it dealt with more physical methods such as hiding the information inside password-protected case. Nevertheless, they served the purpose. As times followed, more concise methods were followed which dealt with obscuring the data itself. This marked the era of digital steganography.

Generally, steganography should meet the few imperative requirements as described below:

1. Imperceptibility to a Human Visual System:
Whenever an image is subjected to watermarking process, its watermarked version should look similar to the one which is not watermarked. There should not be any palpable distinctions so that third party person might learn about watermarking. If this condition is fulfilled then it serves the primary purpose of watermarking.

2. Robust to various kinds of distortions
The watermarked image should be robust and sturdy against distortions such as lossy compressions or any types of modifications. It should produce the original image after reversal of processes. This makes sure that the message to be hidden remains safe even if it gets attacked and manipulated.

3. Simplicity of detection and extraction
For an individual who possesses the private key to retrieve the message, it should be easy to extract it. For others who don't have a key, it should be very complicated to unlock its contents. This ensures that the watermarking is perfect and can be only deciphered by rightful individual.

4. High information capacity
The watermarked image should be able to carry large information without burdening the channel or the original image. This property describes how much data should be embedded for proper carriage.

In this paper, Karhunen-Loève Transform (KLT) has been employed for performing steganography. In order to acquire better image quality, sub-matrix of specific size of original pixel matrix message has been used. Readjustment will be executed to extract secret data exactly and to minimize the perceptual distortion resulted from embedding. The experimental results show that proposed method provides a large embedding capacity, and the quality of the steganographically hidden image is improved as well.

The paper is organized as follows: section II gives idea of proposed solution and section III gives implementation steps. Section IV shows experimental results and section V provides conclusion.

## 2. IDEA OF PROPOSED SOLUTION
In this algorithm the data information are initially compacted utilizing the KLT so as to accomplish a higher hiding limit, then afterward packed into LSBs of carrier image, which is in the RGB spatial domain. This combination will be useful in expanding the limit for hiding large messages, accomplishing a high caliber steganographically hidden image so it is essentially intangible and enhancing the execution time of the algorithm.

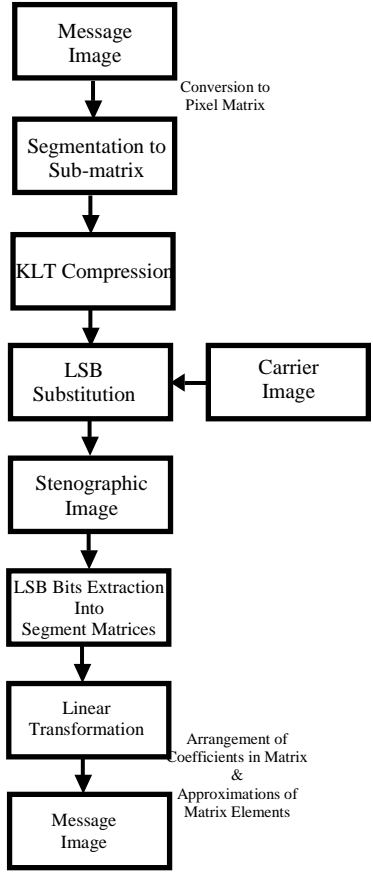

**Fig. 1: Block diagram of proposed algorithm**

## 2.1 Karhunen-Loève Transform

The aim of image compression is to store an image in a more compact form, i.e., a representation that requires lesser bits for encoding than original image. As images have a definite structure, there is some correlation between neighboring pixels. If reversible transformation that removes the redundancy by decorrelating the data, then an image can be stored more effectively. Karhunen-Loève Transform (KLT) is a linear transformation which achieves this goal. The Karhunen-Loève Transform (KLT), also known as the Hotelling Transform, Eigenvector Transform or sometimes Principal Component Analysis is widely considered to achieve optimal signal processing for data representation, compression and analysis [1]-[4]. This transform is used in many domains in which random processes or sequences are encountered.

Consider a set of *n* one or multi-dimensional discrete signals represented as column vectors $f_0, f_1, f_2, .....f_{n-1}$ each having *M* elements. Let us denote the mean vector and covariance matrix of $f_i$ $(i=0,1,2....n-1)$ by $\overline{f}$ and $\sum_f$ respectively. Then the $r^{th}$ element of mean vector is given by

$$\overline{f}(r) = \frac{1}{n}\sum_{i=0}^{n-1} f_i(r) \quad (1)$$

and the $(r,c)^{th}$ element of the covariance matrix is given by

$$\sum_f(r,c) = \frac{1}{n}\sum_{i=0}^{n-1}\{f_i(r) - \overline{f}(r)\}\{f_i(c) - \overline{f}(c)\} \quad (2)$$

Therefore,

$$\overline{f} = \frac{1}{n}\sum_{i=0}^{n-1} f_i \quad (3)$$

and

$$\sum_f(r,c) = \frac{1}{n}\sum_{i=0}^{n-1}(f_i - \overline{f})(f_i - \overline{f})^T \quad (4)$$

Now $\overline{f}$ is *M*-dimensional vector and $\sum_f$ is $M \times M$ matrix. Since $\sum_f$ is a real symmetric matrix, *M* eigenvectors $e_j(j=0,1,2,...M-1)$ of $\sum_f$ corresponding to its eigenvalues $\lambda_j$. Suppose a matrix T with these eigenvectors as its rows, i.e,

$$T = \begin{bmatrix} e_0 \\ e_1 \\ e_2 \\ \vdots \\ e_{M-1} \end{bmatrix} \quad (5)$$

Then it can be written as $T^{-1}=T^T$, and

$$T\sum_f T^T = \begin{bmatrix} \lambda_0 & 0 & 0 & 0 & 0 \\ 0 & \lambda_1 & \cdots & \cdots & 0 \\ \vdots & \vdots & \lambda_2 & \vdots & \vdots \\ \vdots & \vdots & \vdots & \ddots & \vdots \\ 0 & 0 & \cdots & \cdots & \lambda_{M-1} \end{bmatrix} \quad (6)$$

The K-L transform pair is then defined as

$$g_i = T(f_i - \overline{f}) \quad (7)$$

$$f_i = T^T g_i + \overline{f} \quad (8)$$

Since the covariance matrix of transformed vectors is a diagonal matrix, it is evident that the elements of the transformed vector $g_i$ are uncorrelated [5].

## 2.2 Least Significant Bit (LSB) Substitution

LSB substitution method for hiding images is a very simple and easy method to implement. Suppose s is the image to be hidden in host image H. Both are grayscale images with each pixel having n bits. Suppose S is to be embedded in rightmost k-bits of each pixel in H. First, S will be converted to S'. In this process, each pixel of S is decomposed into several small k-bit units to form the k-bit image S'[6].

Consider an 8-bit grayscale bitmap image where each pixel is stored as a byte representing a grayscale value. Suppose the first eight pixels of the original image have the following grayscale values [7]:

10010111 10001100 11010010 01001010 00100110 01000011 00010101 01010111

To hide the letter A whose binary value is 01000001, we would replace the LSBs of these pixels to have the following new grayscale values:

1001011**0** 1000110**1** 1101001**0** 0100101**0** 0010011**0** 0100001**0** 0001010**0** 0101011**1**

As from above example, only half the LSBs need to change. The difference between the host image and the hidden image will be hardly noticeable to the human eye. Figure 1 shows the flowchart for intended operation.

## 2.3 Message Segmentation

The original matrix of size m × n is subdivided into smaller sub-matrix of size s × n. As each pixel is divided into R, G and B pixels, the matrix has increased into the size, three times the original. Hence size of this matrix is 3 m × n. Segregation of larger matrix into smaller matrix reduces time required for intended operation. This makes the process efficient in time domain. Also, reducing matrix size in turn demotes reducing image size, i.e., removing data redundancy. This brings about effect of compression. Thus concealing the secret image can be achieved through this type of compression. This makes procedure efficient in increasing the capacity of image to hide the data.

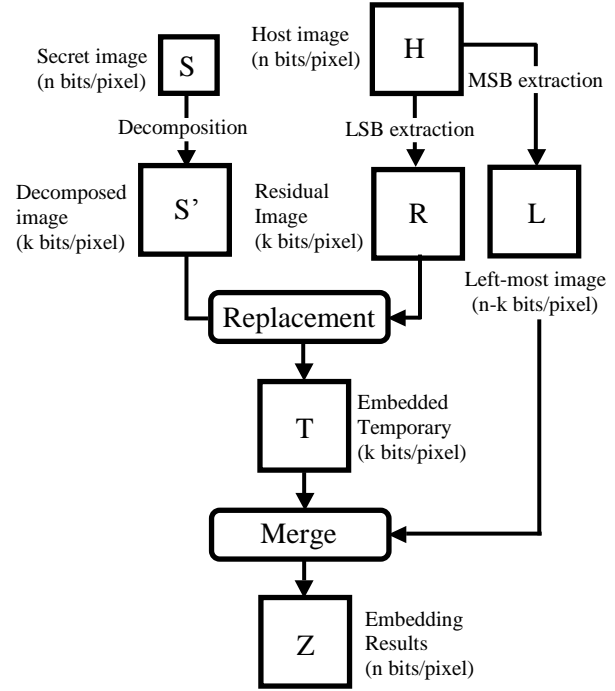

**Fig 2: The flowchart of the embedding process in the simple LSB substitution method**

To segment the image matrix, iterative approach is taken in which classification decision about each pixel can be taken in parallel [8]. Decisions made at neighboring points in the current iteration can be then used to make decision in next iteration. Suppose a set of pixels $\{g_1, g_2, \ldots, g_n\}$ is to be classified into *m* classes $\{C_1, C_2, \ldots, C_3\}$. To keep class assignment of pixels independent, it is assumed that for each pair of class assignment $f_i \in C_j$ and $f_h \in C_k$, there exists a quantitative measure of compatibility $C(i,j;h,k)$ of this pair.

Let $p_{i,j}$ represent the probability that

$$f_i \in C_j, \qquad 1 \le i \le n \qquad (9)$$

$$1 \le j \le m,$$

with $0 \le p_{i,j} \le 1$ and $\sum_j p_{i,j} = 1$.

If $p_{h,k}$ is high and $C(i,j;h,k)$ is positive, $p_{i,j}$ is increased since it is compatible with high probability event $f_h \in C_k$. Similarly, if $p_{h,k}$ is high and $C(i,j;h,k)$ is negative, $p_{i,j}$ is reduced as it is incompatible with $f_h \in C_k$. On the other hand, if $p_{h,k}$ is low or $C(i, j; h, k)$ is nearly zero, $p_{i,j}$ is not changed as either $f_h \in C_k$ has a low probability or is irrelevant to $f_i \in C_j$.

## 3. STEPS OF IMPLEMENTATION

Considering an image as carrier image and another as message image, all the steps were carried out. Carrier image was converted to HSV format initially.

### 3.1 Representation of Secret Message into Corresponding Pixel of Image

The secret message to be hidden is represented as $m \times n$ RGB matrix of pixels, where $m$ represents height of image and $n$ represents width of image. Hence, an image $A$ can be shown as follows:

$$A = \begin{bmatrix} P_{1,1} & P_{1,2} & \cdots & P_{1,n} \\ \vdots & \vdots & & \vdots \\ \vdots & \vdots & \ddots & \vdots \\ P_{m,1} & P_{m,2} & \cdots & P_{m,n} \end{bmatrix} \quad (10)$$

To achieve efficient processing, the pixels were segregated into three distinct pieces of information R,G and B. The matrix $A$ thus becomes of size $3\ m \times n$, which is represented by $A'$.

$$A' = \begin{bmatrix} R_{1,1} & R_{1,2} & \cdots & R_{1,n} \\ G_{1,1} & G_{1,2} & \cdots & G_{1,n} \\ B_{1,1} & B_{1,2} & \cdots & B_{1,n} \\ \vdots & \vdots & \ddots & \vdots \\ R_{m,1} & R_{m,2} & \cdots & R_{m,n} \\ G_{m,1} & G_{m,2} & \cdots & G_{m,n} \\ B_{m,1} & B_{m,2} & \cdots & B_{m,n} \end{bmatrix} \quad (11)$$

### 3.2 Segmentation of Image in Sub-Matrix

The result of image segmentation is a set of segments that collectively cover the entire image, or a set of contours extracted from the image. It separates out foreground and background in meaningful way [9]. Each of the pixels in a region is similar with respect to some characteristic or computed property, such as color, intensity, or texture. Maximum Likelihood Classification (MLC) method is used for this purpose. The iterative clustering algorithm first computes the cluster mean values and covariance matrices, adjusting these values while reading the image data set.

The assignment to the class is based on image statistics. A clustering algorithm groups pixel values with similar statistical properties according to user definition of number of classes [10]. The idea is to identify pixel clouds from the feature space which have similar reflectance values. Each pixel cloud, grouped into clusters, characterizes spectral signature of that object. The cluster information is used to perform spatial assignment of the individual pixels to the derived clusters. The MLC determines to which spectral class each cell in the image has the highest probability of belonging.

### 3.3 Application of KLT

KLT helps in compressing the image matrix further to increase hiding capacity of image. Consider that image has been divided in segments of size s. The matrix A* corresponds to the first of these segments.

$A* =$

$$\begin{bmatrix} x_{1,1} & x_{1,2} & x_{1,3} & \cdots & x_{1,n-1} & x_{1,n} \\ x_{2,1} & x_{2,2} & x_{2,3} & \cdots & x_{2,n-1} & x_{2,n} \\ x_{3,1} & x_{3,2} & x_{3,3} & \cdots & x_{3,n-1} & x_{3,n} \\ \vdots & \vdots & \vdots & \ddots & \vdots & \vdots \\ x_{3s-2,1} & x_{3s-2,2} & x_{3s-2,3} & \cdots & x_{3s-2,n-1} & x_{3s-2,n} \\ x_{3s-1,1} & x_{3s-1,2} & x_{3s-1,3} & \cdots & x_{3s-1,n-1} & x_{3s-1,n} \\ x_{3s,1} & x_{3s,2} & x_{3s,3} & \cdots & x_{3s,n-1} & x_{3s,n} \end{bmatrix}$$

where $x_{3i-2,j} = R_{i,j}$, $x_{3i-1,j} = G_{i,j}$ and $x_{3i,j} = B_{i,j}$ for all $i \in \overline{1,s}$ and $j \in \overline{1,n}$ [11].

The next step is to calculate eigenvalues and eigenvectors of covariance matrix. The distribution of the processed matrix's A* energy (information) among the eigenvectors is indicated by the eigenvalues. Jacobi Eigenvalue Algorithm is used for this purpose. The value of each eigenvalue is proportional to the quantity of energy stored by the corresponding eigenvector. The matrix is same as equation (6).

Since covariance matrix is symmetric, it has orthonormal basis of eigenvectors. These eigenvectors constitute a (3s)×(3s) orthogonal matrix $V = \begin{bmatrix} v_1 & v_2 & \cdots & v_{3s} \end{bmatrix}$, which has the property

$$V \cdot V^T = V^T \cdot V = I_{3s} \text{ [12]} \quad (12)$$

### 3.4 Concealing the Message

The entries of matrix A* are integer and hence can be accommodated in single byte. But this is not the case with entries of projection matrix as the values are real. Hence they need at least 4 bytes to store. In terms of computer bytes, the compression rate would be $\dfrac{4k}{3s}$, providing no compression. To achieve original compression rate, a linear transform is applied on $P_j$ values as follows.

$$P_j{'}_{i,j} = (P_{j_{i,j}} - \min P_j)/(\max P_j - \min P_j) \cdot 255 \tag{13}$$

The similar problem is encountered with matrix V*, the entries are real which need at least 4 bytes to be stored. To round them, following formula can be used:

$$V*'_{i,j} = V*_{i,j} \cdot 32767 \tag{14}$$

After this, information can be hidden using LSB substitution. The number of bits used for hiding will decide the capacity of image to hide the data. Depending on how many of these bits we use for hiding the message, we get three versions of the same algorithm: one bit, two bits and three bits. The more bits we use for hiding, the more information we will be able to hide, at the expense of a greater Carrier Error.

## 3.5 Extraction of Message

The extraction of hidden message is exact reverse process of above procedures. First, the hidden information is extracted from LSBs of steganographically hidden image. For each segment, we get the linearly processed projection and eigenvector matrix for each segment and their dimensions. The next step is to reverse the process of linear transformation using following formulas:

$$aP_{j_{i,j}} = \frac{P_j{'}_{i,j} \cdot (\max P_j - \min P_j)}{255 + \min P_j} \tag{15}$$

$$aV*_{i,j} = \frac{V*'_{i,j}}{32767} \tag{16}$$

The coefficient $a$ signifies that resulted projection matrix and reduced eigenvector matrix are just approximations of their original counterparts. These matrices are combined to obtain an approximation of the initial segment sub-matrix A* as follows:

$$aA* = aV* \cdot aP_j \tag{17}$$

Approximating the RHS,

$$aV* \cdot aP_j \cong V* P_j \tag{18}$$

From equation (14),

$$V* \cdot P_j = V* \cdot (V*)^T \cdot A^* \tag{19}$$

from equation (12),

$$V* \cdot (V*)^T \cdot A^* \cong I_{3s} \cdot A* = A* \tag{20}$$

## 4. EXPERIMENTAL RESULTS

To implement the algorithm, three sets of carrier and message images were used. For each set, different values of LSB bits were used. The algorithm was tested on PC with Intel® Core i5 Quad-Core with 4GB RAM.

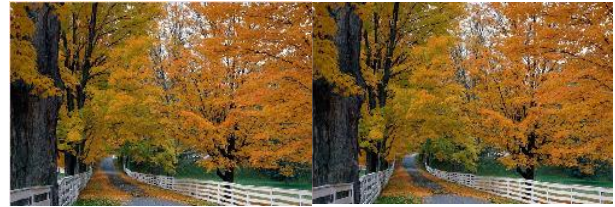

**Carrier Image**     **Steganographically hidden Image**

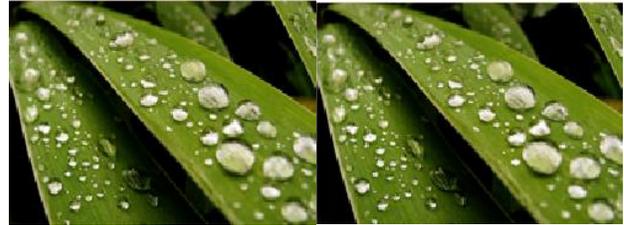

**Message Image**     **Recovered Image**
**Fig.3: Set 1 images**

For the first set of images, one bit was used in LSB. The carrier error is the least 0.395667. It can also be seen from the image that steganographically hidden image is of high quality. The recovered image is too of high quality with message error 1.1348. The hiding time is 3.59876 second and recovery time is 1.63255 seconds.

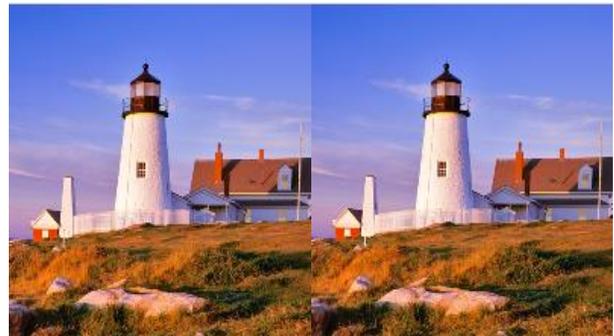

**Carrier Image**     **Steganographically hidden Image**

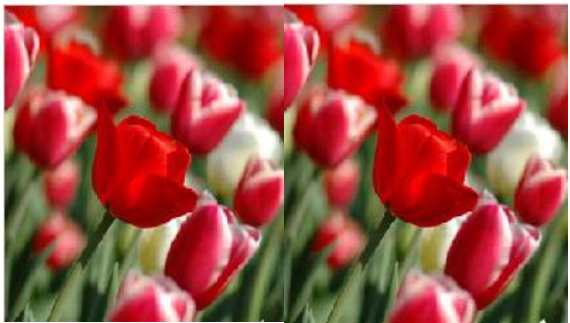

**Message Image**      **Recovered Image**

**Fig.4: Set 2 images**

For the second set of images, two bits were used in LSB. The carrier error is 0.638335, almost double of carrier error of set 1. steganographically hidden image is of average quality. The recovered image has message error 2.50727. The hiding time is 3.10416 seconds and recovery time is 1.317215 seconds.

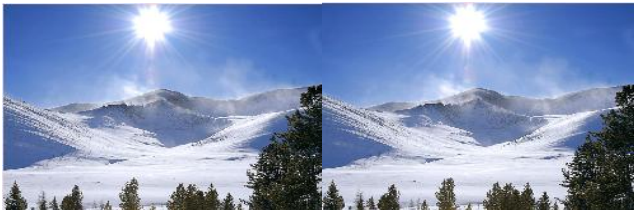

Carrier Image      Steganographically hidden Image

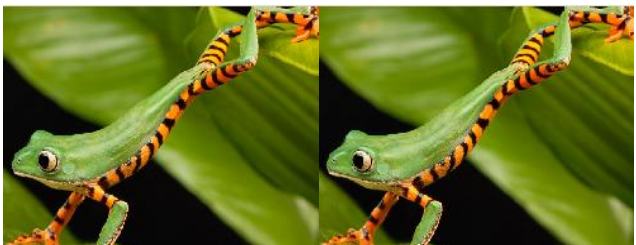

Message Image      Recovered Image

**Fig.5: Set 3 images**

For the third set of images, four bits were used in LSB. The carrier error is maximum, 1.76675. Steganographically hidden image is of poorest quality. The recovered image has message error 1.61832. The hiding time is 7.41073 seconds as the image size was larger and recovery time is 2.59556 seconds.

**Table 1: Parameters calculated during processing**

| Image | Data can be embedded | Compression Rate | Hiding Time | Recover Time | Carrier Error | Message Error |
|---|---|---|---|---|---|---|
| Set1 | 25311 | 0.493827 | 3.59876 | 1.63255 | 0.395667 | 1.1348 |
| Set2 | 9125 | 0.333333 | 3.10416 | 1.31721 | 0.638335 | 2.50727 |
| Set3 | 15160 | 0.444444 | 7.41073 | 2.59556 | 1.76675 | 1.61832 |

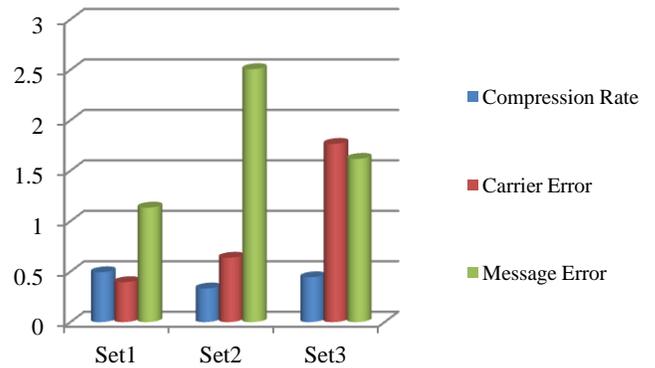

**Fig.6: Graphical representation of parameters**

Table 1 shows different parameters calculated for all three sets of images. It can be seen that carrier error and message error is maximum for third set, i.e., when four LSB bits were used for encoding. As number of bits used for encoding increases, the resemblance of steganographically hidden image with message image starts increasing. Whereas only one bit used for encoding provides no resemblance meaning proper masking. It is evident from table 1 that value of carrier error and message error is the least for this situation.

Fig.6 shows graphical representation of compression rate, carrier error and message error. Above results can be verified from this graph.

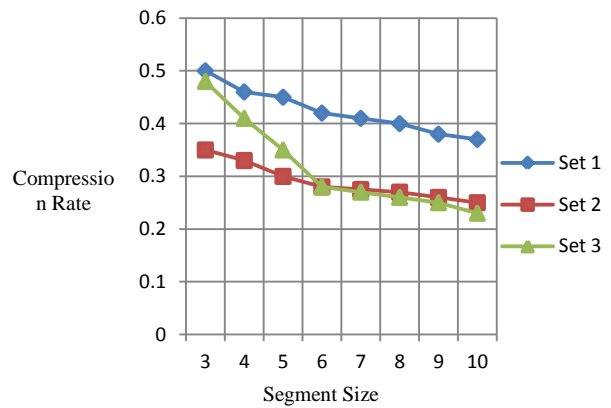

**Fig.7: Plot of segment size vs. compression rate for three sets**

Fig.7 shows plot of segment size vs. compression rates. It can be observed that rate depends largely on number of segments. As number of segments increases, compression rate decreases. The compression rate is poor for less segmented images, implying that compressed message will not "fit" inside the carrier [13].

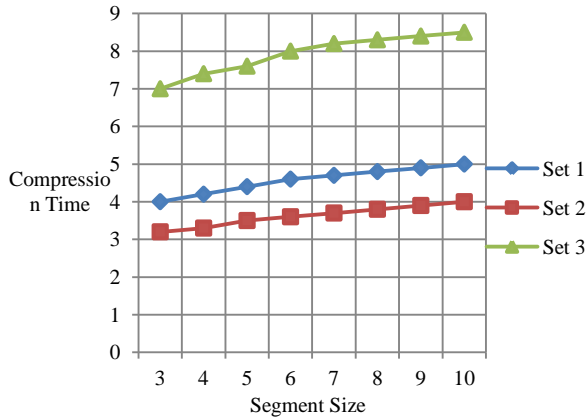

**Fig.8: Plot of segment size vs. compression time for three sets**

Fig.8 shows plot of segment size vs. compression time. It is clear that as number of segments increases, time required will also more. Smaller is the size of segments, faster will be compression meaning lesser time will be required.

## 5. FUTURE WORK
The method can be extended to hiding of various media inside image such as pure text, audio file etc. The capacity of data hiding can be increased more so that multimedia files can be hidden. The algorithm can be made stronger by reviewing recent stego-systems and considering how they can be attacked.

Additional steganalytic functions can be used so that the system is prepared to operate on a wider range of images. This could be done by implementing some more breaching techniques.

## .6. CONCLUSION
The primary aim of this paper was to implement a robust and secure method for steganography which would offer optimum performance. All the three systems have been simulated on three different sets of carrier and message images. From the parameters it was found that encoding with two bits for set 2 type images offers best results as size of carrier image and data hiding capacity of provided are exactly equal. Other systems also provide optimum results with carrier and message errors. By virtue of KLT, the compression is high which adds up to capacity of image to hide the data. It also satisfies the basic rules of data concealing that hidden image should not be detectable. From the results it can be seen that it fulfils this requirement. Thus it can be concluded that algorithm gives a proper way to hide the data with minimum errors and optimum image quality.